# Basic design and engineering of normal-conducting, iron-dominated electromagnets

*Th. Zickler*
CERN, Geneva, Switzerland

**Abstract**
The intention of this course is to provide guidance and tools necessary to carry out an analytical design of a simple accelerator magnet. Basic concepts and magnet types will be explained as well as important aspects which should be considered before starting the actual design phase. The central part of this course is dedicated to describing how to develop a basic magnet design. Subjects like the layout of the magnetic circuit, the excitation coils, and the cooling circuits will be discussed. A short introduction to materials for the yoke and coil construction and a brief summary about cost estimates for magnets will complete this topic.

## 1    Introduction

The scope of these lectures is to give an overview of electromagnetic technology as used in and around particle accelerators considering *normal-conducting, iron-dominated* electromagnets generally restricted to direct current situations where we assume that the voltages generated by the change of flux and possible resulting eddy currents are negligible. Permanent and superconducting magnet technologies as well as special magnets like kickers and septa are not covered in this paper; they were part of dedicated special lectures.

It is clear that it is difficult to give a complete and exhaustive summary of magnet design since there are many different magnet types and designs; in principle the design of a magnet is limited only by the laws of physics and the imagination of the magnet designer. Furthermore, each laboratory and each magnet designer or engineer has his own style of approaching a particular magnet design. Nevertheless, I have tried to gather general and common principles and design approaches.

I have deliberately focused on applied and practical design aspects with the main goal of providing a guide-book with practical instructions on how to start with the design of a standard accelerator magnet. As far as I know, there is no such manual that provides step-by-step instructions allowing the setting up of a first, rough analytic design before going into a more detailed numerical design with field computation codes like ROXIE, OPERA, ANSYS, or POISSON. This guide-book should also help to assess and validate the feasibility of a design proposal and to draft a list of the key parameters (with just pencil and paper) without spending time on complex computer programs.

Please keep in mind that these lectures are meant for students of magnet design and engineering working in the field of accelerator science — not for advanced experts.

For the sake of briefness and simplicity I have refrained from deriving once again Maxwell's equations — they have been extensively treated by experts in other lectures. You will also find mathematics reduced to a bare minimum. The derivation of formulas in this text might sometimes appear condensed, but in case you want to learn more, you should always be able to find the sources with the help of the bibliography cited at the end. To guarantee consistency throughout, SI (MKSA) units are used systematically.

The paper starts with a short introduction to basic concepts and magnet types, followed by a section dedicated to collecting information and defining the requirements and constraints before starting the actual design. The main part gives an introduction to basic analytic magnet design covering topics such as yoke design, coil dimensioning, cooling layout, material selection, and cost estimation. Although the lectures presented during the course included a section introducing numerical design methods, it has been omitted from the proceedings since it was found to be too exhaustive. The bibliography recommends literature for further reading for those who wish to go more deeply into this subject.

## 2    Basic concepts and magnet types

We introduce basic concepts and classical normal-conducting magnet types, highlight their main characteristics, and explain very briefly their function and purpose in a particle accelerator.

### 2.1    Dipoles

In a circular particle accelerator or in a curved beam transfer line, dipoles are the most commonly used elements. A dipole provides a uniform field between its two poles which is excited by a current circulating in the coils. The system follows the right-hand convention, i.e., a current circulating clockwise around the poles produces a magnetic field pointing downwards.

Their purpose is to bend or steer a charged particle beam. Applying again the right-hand rule, when a beam of positively charged particles directed into the plane of the paper sees a field pointing downwards, it is deflected to the left, as shown in Fig. 1 (a).

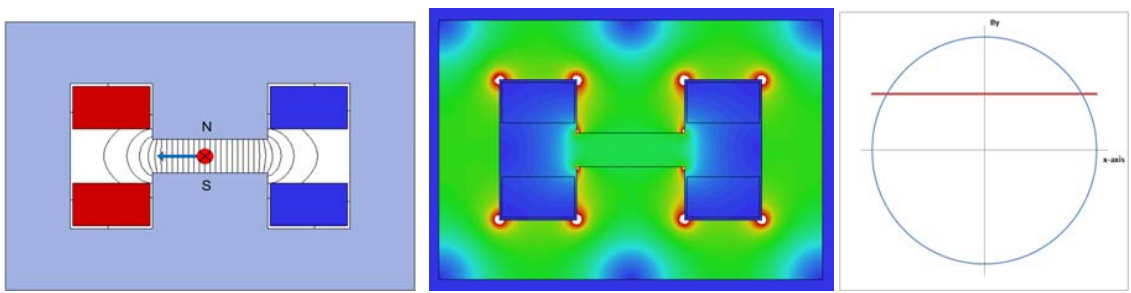

**Fig. 1:** Dipole: cross-section (a), 2D-field distribution (b), and field distribution on the *x*-axis (c)

The equation describing a normal ideal (infinite) pole is:

$$y = \pm r$$

where *r* is the half-gap height. The magnetic flux density between these two poles is ideally constant and has only a component in the *y*-direction, as one can see from Fig. 1 (a)–(c):

$$B_y = a_1 = B_0 = \text{const.}$$

In an ideal dipole only harmonics of: $n = 1$, 3, 5, 7... (= 2$n$ pole errors) can appear. These are called the 'allowed' harmonics.

### 2.2    Quadrupoles

The second most commonly used magnetic elements are quadrupoles. Their purpose is to focus the beam. Note that a horizontally focused beam is at the same time vertically defocused. A quadrupole has four iron poles with hyperbolic contour which can be ideally described for a normal (non-skew) quadrupole by

$$2xy = \pm r^2 \; ,$$

where $r$ is the aperture radius.

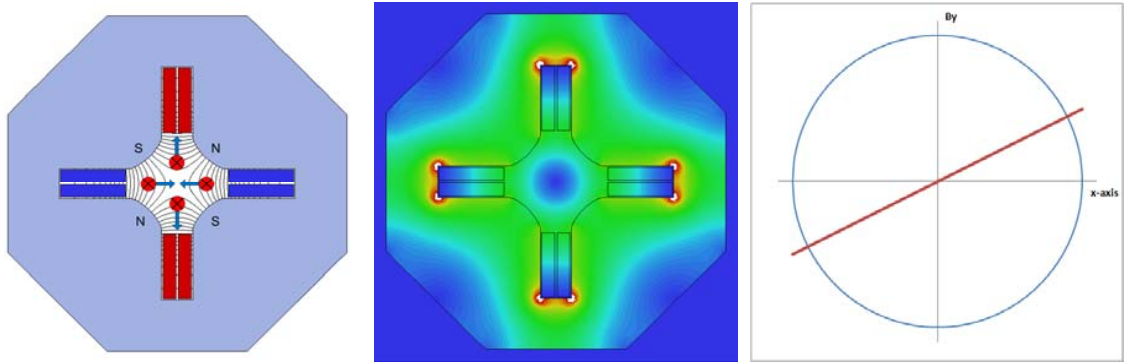

**Fig. 2:** Normal quadrupole: cross-section (a), 2D-field distribution (b), and field distribution on the $x$-axis (c)

A quadrupole provides a field which is zero at the centre and increases linearly with distance from the centre as shown in Fig. 2 (c). The equipotential lines are hyperbolas ($xy$ = const.) and the field lines are perpendicular to them. Dipoles and quadrupoles are linear elements, which means that the horizontal and the vertical betatron oscillations are completely decoupled. The Cartesian components of the flux density in an ideal quadrupole are not coupled; the $x$-component in a certain point only depends on the $y$-coordinate and the $y$-component only depends on the $x$-coordinate following the relation

$$B_y = a_2 x \quad \text{and} \quad B_x = a_2 y \; .$$

With the polarity shown in Fig. 2 (a), the horizontal component of the Lorentz force on a positively charged particle moving into the plane of the drawing, is directed towards the axis; the vertical component is directed away from the axis. This case thus exhibits horizontal focusing and vertical defocusing.

The 'allowed' harmonics in an ideal quadrupole are: $n = 2$, 6, 10, 14, ... (= $2n$ pole errors).

## 2.3 Sextupoles

Sextupoles can be found in circular accelerators and less often in transfer lines. They have six poles of round or flat shape. Their main purpose is to correct chromatic aberrations: particles which are off-momentum will be incorrectly focused in the quadrupoles, which means that high-momentum particles with stronger beam rigidity will be under-focused, so that betatron oscillation frequencies will be modified. A positive sextupole field can correct this effect and can reduce the chromaticity to zero, because off-momentum particles circulate with a radial displacement with respect to the ideal trajectory and see therefore a correcting field in the sextupole as shown in Fig. 3 (a). We have also seen that the first 'allowed' harmonic in a dipole is the sextupole component, which leads to a resulting negative chromaticity requiring compensation by distinct sextupole elements.

The equation for a normal (non-skew) sextupole with ideal poles is

$$3x^2 y - y^3 = \pm r^3$$

where $r$ is again the aperture radius. The magnetic field varies quadratically with the distance from the magnet centre as one can see in Fig. 3 (c). Sextupoles are non-linear elements, which means that the y-component of the flux density at a certain point in the aperture depends on both the $x$- and $y$-coordinate, and is described by

$$B_y = a_2(x^2 - y^2)$$

The 'allowed' harmonics in an ideal sextupole are: $n = 3, 9, 15, 21...$ (= $2n$ pole errors).

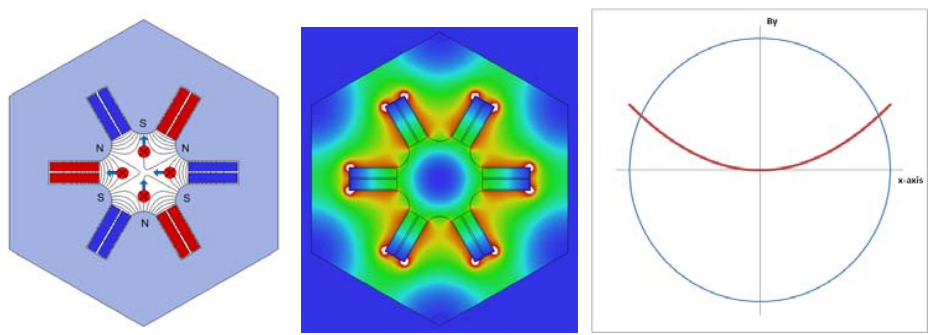

**Fig. 3:** Normal sextupole: cross-section (a), 2D-field distribution (b), and field distribution on the $x$-axis (c)

## 2.4    Octupoles

Octupoles are quite rarely used and can be mainly found in colliders and storage rings. Amongst other purposes, they are used for 'Landau' damping, to introduce a tune-spread as a function of the betatron amplitude, to de-cohere the betatron oscillations, and to reduce non-linear coupling. The eight poles of a normal (non-skew) ideal octupole as shown in Fig. 4 (a) and (b) follow the equation

$$4(x^3y - xy^2) = \pm r^4$$

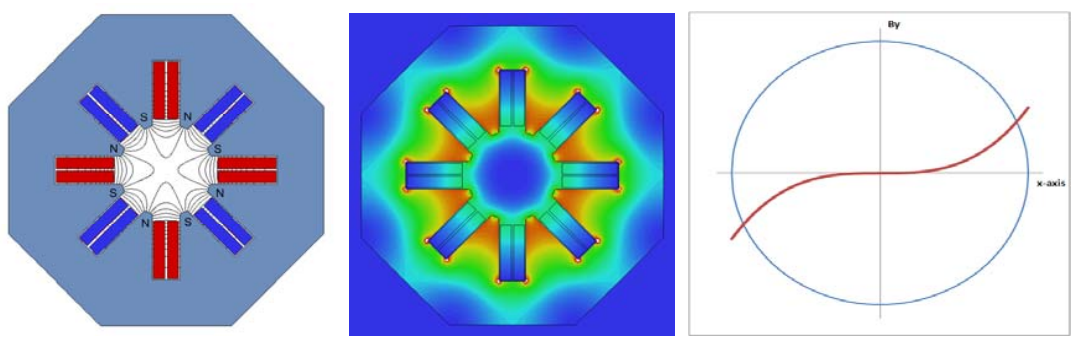

**Fig. 4:** Normal octupole: cross-section (a), 2D-field distribution (b), and field distribution on the $x$-axis (c)

The $y$-component of the magnetic flux density in any point of the aperture can be described by the following relation

$$B_y = a_4(x^3 - 3xy^2)$$

The 'allowed' harmonics in an ideal octupole are: $n = 4, 12, 20, 28...$ (= $2n$ pole errors).

## 2.5 Skew magnets

Skew versions exist for all of the above-described magnet types. Skew means a rotation of the magnet along the longitudinal axis by $90°/n$, where $n$ is the index of the main field component (i.e., $n = 1$ for dipole, $n = 2$ for quadrupole, $n = 3$ for sextupole). Rotating linear magnetic elements leads to loss of the betatron decoupling. Fig. **5** shows a skew quadrupole, the purpose of which is to control the coupling of horizontal and vertical betatron oscillations. In a skew quadrupole, a beam that is displaced in the horizontal plane is deflected vertically, and a beam that is displaced in the vertical plane is deflected horizontally.

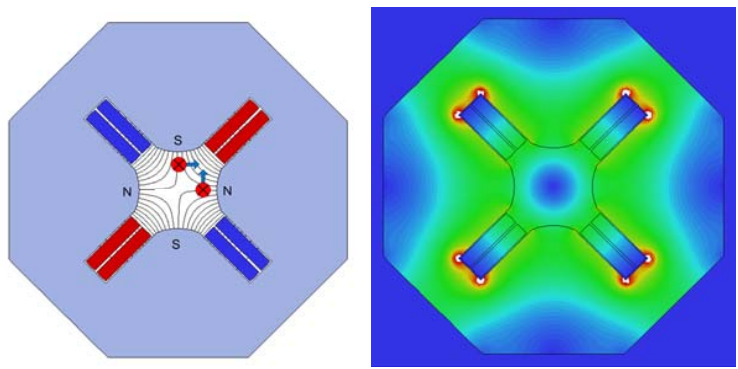

**Fig. 5:** Skew quadrupole: cross-section (a), 2D-field distribution (b)

## 2.6 Combined-function magnets

Combined-function magnets unite several main field components in one magnet, e.g., a dipole and a quadrupole. We can distinguish between two types of combined-function magnet.

These are magnets where the different functions are generated by the sum of scalar potentials and the shape of the pole, and magnets where the different functions are generated by separate coils individually powered.

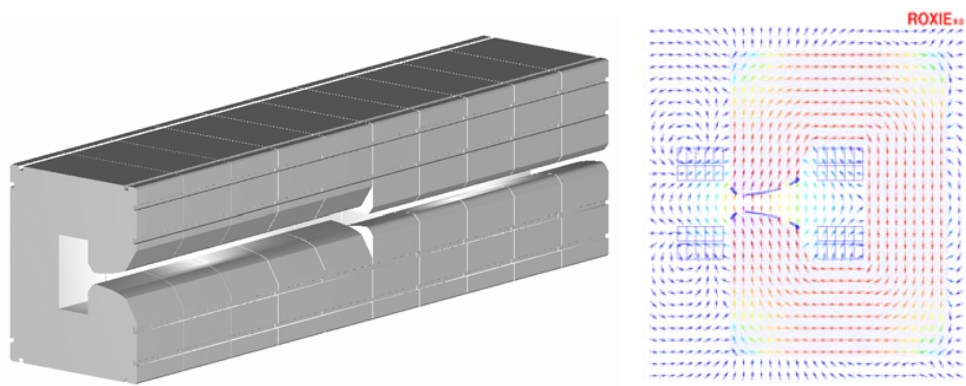

**Fig. 6:** Combined-function magnet yoke of the CERN Proton Synchrotron

The second type is of minor importance and sometimes used when limited space in the machine demands special solutions. An example of a quadrupole with integrated steering coils is illustrated in Fig. 7. Other types combine sextupoles with steering functions or quadrupoles with sextupoles. The advantage here is that the amplitudes of both field components can be adjusted independently, but often the field quality of one function is significantly reduced. In the example shown, the dipole field suffers from a strong sextupole component because the yoke geometry has been tailored to develop a quadrupole field.

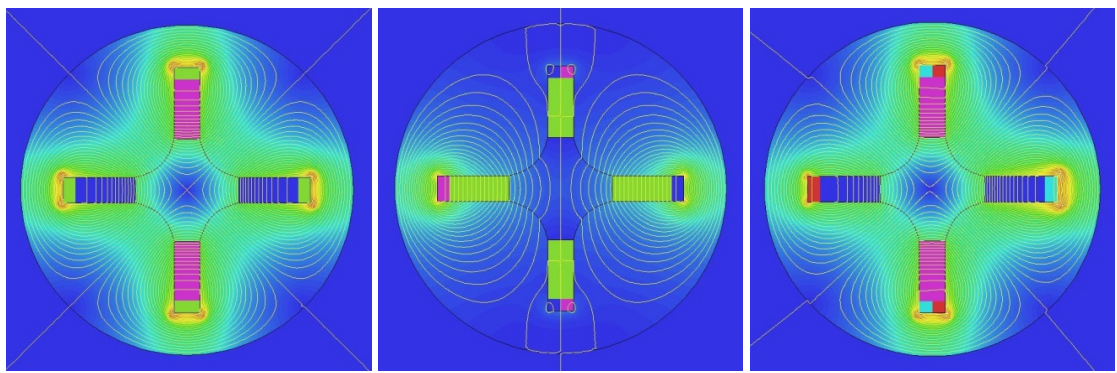

**Fig. 7:** Quadrupole with integrated steering coils: quadrupole field only (a), dipole field only (b), dipole field superimposed on quadrupole (c)

A solution to circumvent this field quality problem is shown in Fig. 8, where several functions (horizontal dipole, vertical dipole, quadrupole, skew quadrupole, and sextupole) are incorporated in one magnet. The field distribution in this case is solely determined by the conductor geometry and not by iron poles. In Fig. 8 (a) only the coils providing the horizontal dipole field are powered, while Fig. **8** (b) illustrates the field distribution when all magnetic functions are excited.

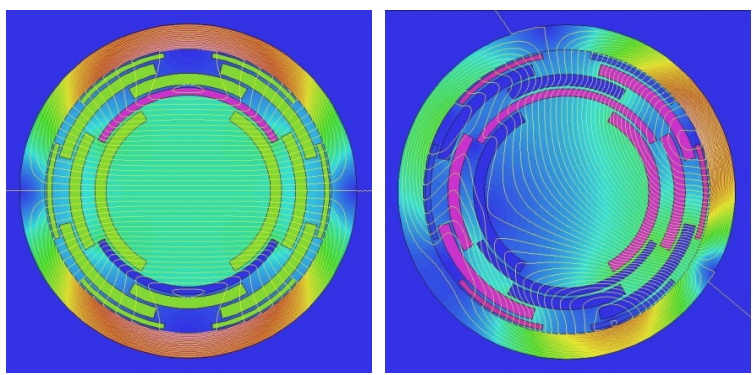

**Fig. 8:** Nested combined-function corrector: vertical dipole (a) and combined field distribution (b)

## 2.7 Solenoids

Although solenoids are strictly speaking not iron-dominated magnets, they will be briefly introduced here for the sake of completeness. A lecture entirely dedicated to solenoids can be found in these proceedings.

Solenoids are relatively simple lenses with a field created by a rotationally symmetric coil. From Maxwell's equation $\mathrm{div}B = 0$, the magnetic field, which is purely longitudinal in the inner part of the coil, must contain radial components at the entrance and at the exit. While particles moving exactly on the axis do not experience any force, the others suffer an azimuthal acceleration due to the radial component while entering and leaving the lens. Because of the azimuthal motion there is a radial force in the longitudinal field. This force is proportional to the radial distance from the axis. To increase the field close to the axis and to capture and limit the stray field, solenoid coils are usually surrounded by an iron yoke.

## 3    What do we need to know before starting?

Before one can enter into the design of a magnet all relevant information which will have an influence on the design, construction, installation, and operation of the future magnet has to be put together. What the term 'relevant' means is explained in this section preceded by a brief discussion about goals in magnet design and magnet life cycles.

### 3.1    Goals in magnet design

We should always keep in mind that the goal in magnet design is to produce a device which is just *good enough* to perform *reliably* with a sufficient *safety factor* at the *lowest cost* and *on time* [1].

What 'on time' means should be obvious: in particular in commercial projects, a delay in the start-up of the operation will result in financial losses. The meaning of 'lowest cost' should also be clear. We will see later how the costs can be optimized. But what does 'good enough' mean? On each project, the obvious parameters such as magnetic field, magnet aperture, magnet dimensions, power consumption, etc. are more or less clearly specified, but it is the tolerances on these parameters that are very often challenging to define. They are a function of the expected machine performance and acceptable deviations from an ideal machine. In this context, orbit distortions, dynamic aperture, tune width, and transfer efficiency could be mentioned, which can be calculated analytically, but nowadays this is usually done numerically. Nevertheless, the interpretation of the results is not straightforward and in many cases the tolerances which are requested by the accelerator physicists tend to be unnecessarily tight. Overly tight tolerances lead to increased costs. An enhanced communication between the magnet designer and the accelerator physicist and mutual understanding can help to solve this problem.

The term 'reliably' basically means to get the Mean Time Between Failures (MTBF) and the Mean Time To Repair (MTTR) to a reasonably low level. Probability theory and risk analysis are well established for industrial engineering and more and more applied now by physicists working in an experimental environment. But for a new design the reliability is usually unknown so one counts on the experience of the magnet engineer to search for a compromise between extreme caution and extreme risk. A detailed design analysis in the framework of an expert review can be helpful in finding this well balanced compromise before proceeding with magnet manufacture.

The last term to be considered is the 'safety factor'. In many projects, the initial design parameters were raised after a few years of operation. Applying a safety factor allows operating a device under more demanding conditions than those initially foreseen but it also permits operating under nominal conditions with less wear, and design flaws are less critical. Since safety factors are typically linked to a rise of production costs, they need to be negotiated between the project engineer and the management. However, the pileup of arbitrary and redundant safety factors at multiple project levels has to be avoided because it leads to an unnecessary increase of costs.

### 3.2    Magnet life cycle

The flow diagram in Fig. 9 shows the typical life cycle of a magnet from the design and construction to the installation and operation and to its final disposal or destruction. We will concentrate mainly on the part which is related to design and calculation. This phase can be split up into different steps which are followed more or less sequentially with possible feedback loops at certain stages. At the beginning of each project the requirements, constraints, and boundaries have to be defined. From this set of parameters a first analytic design should be derived followed by a basic numerical design. After each of the sequential steps (electrical design, mechanical design, integration assessment, and cost estimation) one or more re-iterations of the analytic design might become necessary. Once these steps deliver satisfactory results, an advanced numerical design including field optimizations can be launched.

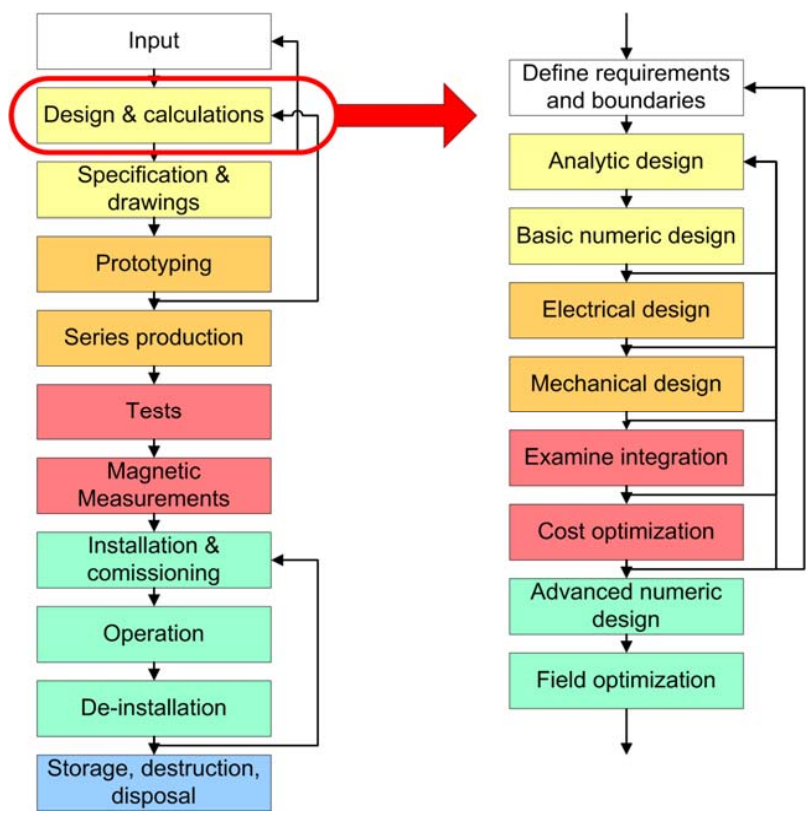

**Fig. 9:** Magnet life cycle

### 3.3 Input parameters

It is essential to realize that a magnet is not a stand-alone device. Throughout its life, a magnet has various interactions with other devices and services. These interactions have to be fully considered in the design phase and the magnet designer calls on his experience to ensure that nothing is forgotten. Ignoring one of the key aspects may result in implementing difficult modifications on the finished product. The main interaction partners are summarized in Fig. 10. Some of them like beam optics, power converters, and cooling are obvious and therefore always taken into account from the beginning. Others, such as vacuum, survey, and integration are often considered in a later stage of the project but sometimes too late, thus complicating the life of the involved parties unnecessarily. Examples of partners which are most likely to be forgotten are safety and transport, with the result that substantial and expensive engineering modification might become necessary in order to install or operate a magnet safely. A good and regular communication with all potential partners from the very start of the project and a clear definition of the interfaces can help to avoid such issues.

It is good practice to contact the responsible partners, collect all necessary information, understand the requirements, constraints and interfaces, and summarize them in a functional specification to be finally approved by each of the involved parties before starting the actual design work.

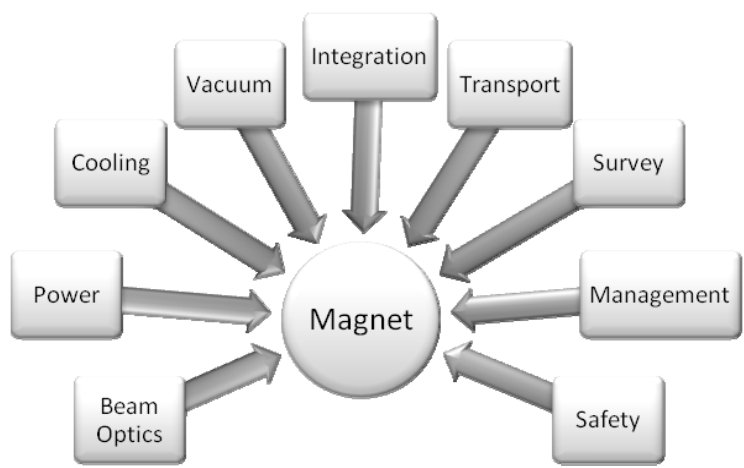

**Fig. 10:** Magnet interaction partners

The following paragraph should give a general idea of possible requirements and help to set up a check-list of information a magnet designer might need to bring together for creating a comprehensive magnet design.

### 3.3.1 General requirements

First of all the magnet type (dipole, quadrupole, sextupole, octupole, combined-function, solenoid, etc.) and its main purpose (bending, steering, charge stage separation, etc.) need to be defined. It is also important to know where the magnet is foreseen to be installed. It makes a difference for the performance of a magnet whether it will be installed in a storage ring, a synchrotron light source, a collider, a pre-accelerator, or in a beam transport line. The tolerances on storage ring magnets are generally more demanding than on accelerators, because the phase spaces of the beams have to be maintained for many revolutions. It also makes sense to discuss the spares policy with the project management at this stage. To foresee a certain number (typically 10%) of spare units and manufacture them together with the units to be installed helps to reduce or avoid down-time in the case of a magnet failure and to save money at the same time. Spare magnets which are produced afterwards or in a case of urgency are inevitable much more expensive.

### 3.3.2 Performance requirements

To start with the initial design work, a magnet designer needs to know at least the basic performance parameters, which are typically provided by the accelerator physicists:

- Beam parameters: type of beam (mass and charge state), energy range and deflection angle ($k$-value in case of quadrupoles).

- Magnetic field: integrated field (or integrated gradient in case of quadrupoles); alternatively the local field (gradient) and magnetic length can be defined.

- Aperture: physical (mechanical) aperture and useful magnetic aperture ('good field region').

- Operation mode: continuous operation, pulsed-to pulse modulation, fast pulsed, definition of the magnetic cycle (an example is shown in Fig. 11) and ramp rates.

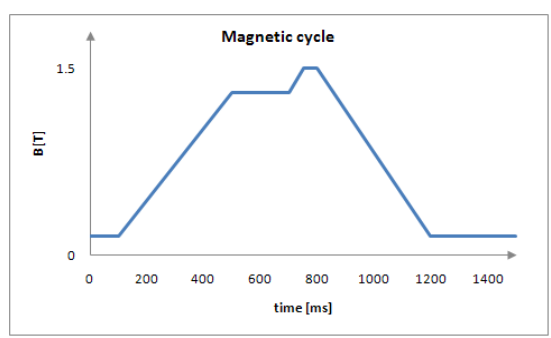

**Fig. 11:** Magnetic cycle

- Field quality: requirements on field homogeneity (uniformity), the allowed harmonic content, requirements on stability and reproducibility, maximum settling time (time constant) for transient effects generated by eddy currents. A simple but lucid method to judge the field quality of a magnet is to plot the homogeneity of the field or the gradient along the boundary of the defined good field region. Achieving the following homogeneity values is reasonable but nevertheless challenging.

Dipole:
$$\frac{\Delta B}{B_0} = \frac{B_y(x,y) - B_y(0,0)}{B_y(0,0)} \leq 0.01\%$$

Quadrupole:
$$\frac{\Delta B'}{B'_0} = \frac{B'_y(x,y) - B'_y(0,0)}{B'_y(0,0)} \leq 0.1\%$$

Sextupole:
$$\frac{\Delta B''}{B''_0} = \frac{B''_y(x,y) - B''_y(0,0)}{B''_y(0,0)} \leq 1\%$$

### 3.3.3    Physical requirements

Important for the mechanical layout — which is of course always linked to the magnetic layout — is to indentify whether geometric boundaries or constraints exist. This can be either a limit in the available space in the accelerator or the beam line, a transport limitation like the maximum allowed charge of an existing crane or a weight limitation of the supporting ground. In particular, the accessibility of the installed magnet should be mentioned here. A magnet designer has not only to assure that the magnet can be transported to its position in the machine, but he has also to take care that sufficient space around the magnet is available to handle the electrical and hydraulic connections and to allow unrestricted access to the reference targets so that the survey group can align the magnet accurately in its final position.

### 3.3.4    Interfaces

The interaction of the magnet with other equipment like power converters, cooling infrastructure, and vacuum systems is quite obvious, but is nevertheless repeated here, since a clear communication and a mutual understanding between the involved groups is essential to avoid any misinterpretation or oversight. It is important to make contact with these groups in an early phase of the design process to clearly define the interfaces and to avoid developing equipment in diverging directions. In this context an example would be a fast-pulsed power converter that cannot be matched to the inductance of the magnet. Another example would be a UH vacuum system requiring in situ bake-out, but the magnet aperture does not then allow installion of such equipment.

The interfaces to the following equipment must be unambiguously defined:

– Power converter: maximum current, maximum voltage, operation mode (pulsed or dc), maximum RMS current, requirements on stability and reproducibility (minimum current), possible control strategies (feed-back, feed-forward), maximum current ramp rates.

– Cooling: available cooling power, maximum flow rate, maximum available pressure, water quality (see Section 4.6.6), circuit type (aluminium or copper), water inlet temperature, temperature stability.

– Vacuum system: size of vacuum chamber, wall thickness and material properties to evaluate potential eddy current issues (only for pulsed magnets), required space for pumping ports and bake-out system, a captive vacuum chamber requiring opening the magnet.

Apart from the above-mentioned points, a magnet may interact with other equipment related to beam diagnostics and monitoring, injection and extraction, RF and control, to name just a few.

### 3.3.5    *Environmental aspects*

Since environmental aspects are not always evidently related to the performance of a magnet, they often don't get the deserved and necessary attention. Consequently, designers and engineers have to be explicitly asked to take care that these aspects are considered and respected in the magnet design. Neglecting such aspects can lead to serious performance problems with the magnet or surrounding equipment. Remedies for such problems are often complicated and costly. In some rare case where it is impossible to find a suitable solution this negligence can even put the whole project into jeopardy.

Since this field covers a wide spectrum and depends on many parameters, it is impossible to present a universal and exhaustive list of all potential risks, eventualities and dangers. The focus is put on the most common issues, but it has to be understood that it is the clear responsibility of a magnet designer or engineer to look beyond the issues stated here, to critically analyse the environment and to identify and point out all possible risks which could endanger the correct performance of the magnet or the surrounding environment. It can be helpful for such an analysis to bear in mind that the interactions are often bi-directional. This means that the magnet can have an influence on the environment, but the environment can also have an influence on the magnet. The following examples serve to illustrate this principle:

*Temperature*: elevated environmental temperatures can influence the dew-point such that condensation appears on the surface of water-cooled coils. On the other hand the amount of heat dissipated from the magnet into the tunnel can be so large that it exceeds the capability of the ventilation system and cannot be removed from the tunnel, causing the temperature to rise.

*Ionizing radiation*: is a specific subject that requires special attention (and which is beyond the scope of this lecture). I would just like to mention the need to select radiation-hard materials and components for accelerator magnets exposed to high radiation levels. The operation of magnets in such an environment also calls for a dedicated design allowing fast repair or replacement, in order to reduce the human intervention time to a minimum.

*Electromagnetic compatibility*: magnetic fringe fields emitted from the magnets can disturb nearby equipment, such as sensitive beam diagnostic devices, while surrounding equipment made of magnetic material can divert part of the magnetic flux and so locally perturb the field quality.

*Safety aspects also have to be seen in this bi-directional way*: covers protect the magnet from effects of the environment (dust, accidental water contact), but they also protect the environment from hazards potentially generated by the magnet (electrocution, burning from hot parts).

Table 1 briefly summarizes the aspects which have been discussed in detail above.

**Table 1:** Aspects to be considered by the magnet designer

| General requirements | Magnet type and purpose |
|---|---|
| | Application |
| | Quantity |
| Performance requirements | Beam parameters |
| | Field requirements |
| | Magnet aperture |
| | Operation mode |
| Physical requirements | Geometric boundaries |
| | Accessibility |
| Interfaces | Power converter |
| | Cooling |
| | Vacuum |
| Environmental aspects | Temperature |
| | Ionizing radiation |
| | Electromagnetic compatibility |
| | Safety |

## 4    Basic analytical design

Before entering into an extensive and detailed two- or three-dimensional magnetic field study using various available software packages allowing the calculation of field distribution and field quality of complex magnetic assemblies, a basic analytic and conceptual design is necessary. Such an approach will allow one to derive the most important characteristics and parameters of the future magnet with a relatively good accuracy and help one to find a reasonable starting point for the numerical design work (as well as the optics design) and thus reduce the number of design iterations.

A magnet is an assembly of different components. Fig. **12** shows a typical normal-conducting, iron-dominated electromagnet — in this case a quadrupole — and its main components: the magnetic circuit, the excitation coils, the cooling circuit, the alignment targets, the sensors and interlock devices, electrical and hydraulic connections, and the magnet support.

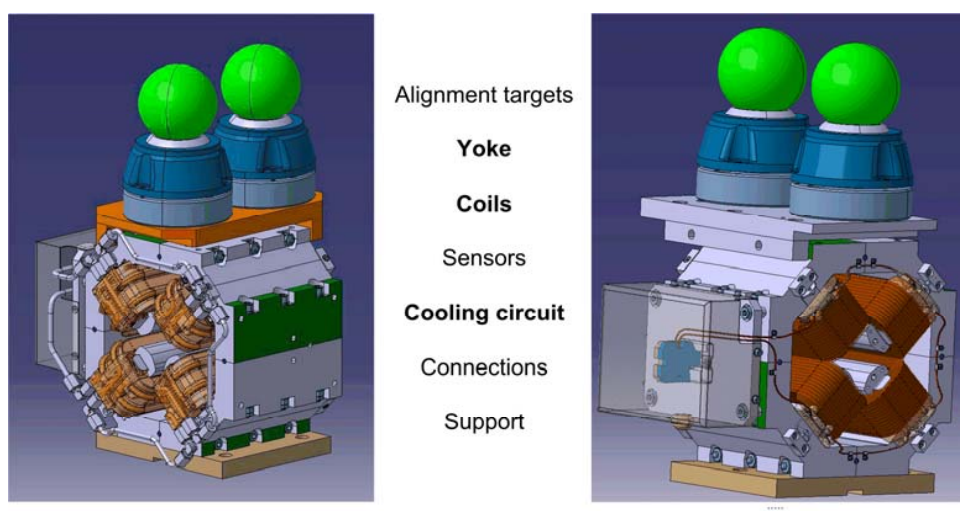

**Fig. 12:** Magnet main components

The following sections will explain how to design the magnetic circuit and the coils and how to dimension the cooling circuits. The other components will not be discussed here since they have no direct influence on the magnet performance and are rather part of the mechanical design.

## 4.1 Dipole yoke design

The first step is to derive the geometry of the magnetic circuit or magnet yoke. This means we have to translate the beam optics requirements into a magnetic design defining the yoke characteristics such as the magnetic induction, the aperture size, and the magnet excitation (ampere-turns).

### 4.1.1 Beam rigidity

A good starting point to define the necessary magnetic induction is to determine the beam rigidity as a function of the particle type and the envisaged beam energy. The beam rigidity $B\rho$ in [Tm] describing the stiffness of a beam can be seen as the resistance of a particle beam against a change of direction when applying a bending force and is defined as

$$B\rho = \frac{1}{qc}\sqrt{T^2 + TE_0}$$

(1)

where $q$ is the particle charge number in [C, coulomb], $c$ is the speed of light in [m/s], $T$ is the kinetic beam energy in [eV], and $E_0$ is the particle rest mass energy in [eV] which is 0.51 MeV for electrons and 938 MeV for protons.

### 4.1.2 Magnetic induction

From the beam rigidity and the assumed bending radius of the magnets we can calculate the flux density or magnetic induction[1] $B$ in [T] for a dipole magnet

$$B = \frac{B\rho}{r_M}$$

(2)

with $r_M$ being the magnet bending radius in [m].

### 4.1.3 Aperture size

The aperture size of a magnet as presented in Fig. 13 is mostly determined by a central region around the theoretical beam trajectory. This region is referred to as 'good field region' and defines the region where the field quality has to be within certain tolerances. The good field region can be circular, rectangular, or elliptical, and takes into account the maximum beam size as well as a certain margin for closed orbit distortions (5–10 mm).

The maximum beam size can be calculated with the help of Eq. (3) which takes into account the lattice functions (beta functions $\beta$ and dispersion $D$), the geometrical transverse emittances $\varepsilon$, which are energy dependent and the momentum spread $\left(\frac{\Delta p}{p}\right)$

---

[1] Generally speaking $B$ has to be a vector. For our purpose it is sufficient and correct to assume that only the main field component in the $y$-direction is present, so $B$ can be written as a scalar. Analogous considerations can be made for quadrupoles and sextupoles.

$$\sigma = \sqrt{\varepsilon\beta + \left(D\frac{\Delta p}{p}\right)^2}$$

(3)

For the beam envelope, a few sigma are typically assumed. The largest beam sizes can usually be expected at injection energy. The total required aperture size is the sum of the good field region, the vacuum chamber thickness (0.3–2 mm) and a margin for installation and alignment (0–5 mm).

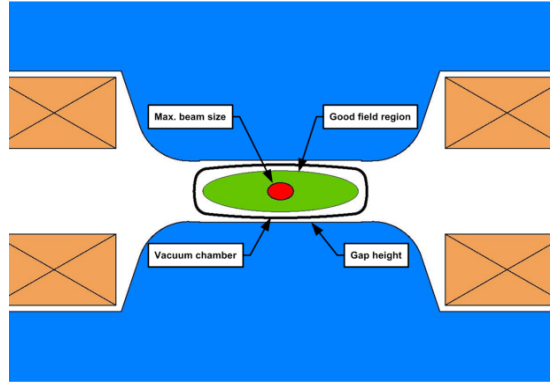

**Fig. 13:** Defining the aperture size

Please note that the numbers in parentheses are typical values for synchrotrons and are meant to give an indication for the order of magnitude. Depending on the individual case they can be significantly different from the quoted numbers.

### 4.1.4 Excitation current

Knowing the aperture of the magnet we can continue to calculate the excitation current in the coils required to drive the desired field strength.

Ampere's law
$$\oint \overrightarrow{H \cdot dl} = NI$$

and
$$\overrightarrow{B} = \mu\overrightarrow{H}$$

and
$$\mu = \mu_0\mu_r$$

leads to
$$NI = \oint\frac{\overrightarrow{B}}{\mu}\cdot\overrightarrow{dl} = \int_{gap}\frac{\overrightarrow{B}}{\mu_{air}}\cdot\overrightarrow{dl} + \int_{yoke}\frac{\overrightarrow{B}}{\mu_{iron}}\cdot\overrightarrow{dl} = \frac{Bh}{\mu_{air}} + \frac{B\lambda}{\mu_{iron}}$$

(4)

when we integrate $B$ along a closed path as shown in Fig. 14 and assume that $B$ remains constant along this path.

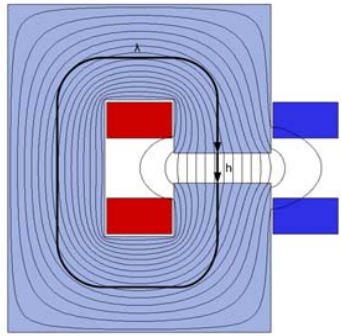

**Fig. 14:** Closed integration path in a dipole magnet

The gap height is indicated by $h$ and the mean flux path in the iron circuit by $\lambda$. As long as the iron is not saturated we can further assume that

$$\frac{h}{\mu_{air}} \gg \frac{\lambda}{\mu_{iron}}$$

such that Eq. (4) can be simplified to

$$NI_{(per\ pole)} = \frac{Bh}{2\eta\mu_0} \qquad (5)$$

where $h$ is the magnet gap height in [m], $H$ is the magnetic field vector in [A/m], $\eta$ is the efficiency (typically 99%), $\mu_0$ is the permeability of free space ($4\pi\ 10^{-7}$ [Vs/Am]), and $\mu_r$ is the relative permeability ($\mu_{air} = 1$ and $\mu_{iron} > 1000$ if not saturated). Note that Eq. (5) is only approximate and neglects fringe fields and iron saturation.

### 4.1.5    Reluctance and efficiency

In analogy to Ohm's law, one can define the 'resistance' of a magnetic circuit, called 'reluctance', as

$$R_M = \frac{NI}{\Phi} = \frac{l_M}{A_M\mu_r\mu_0} \qquad (6)$$

with $\Phi$ indicating the magnetic flux in [Wb], $l_M$ the flux path length in the iron part in [m] and $A_M$ the iron cross-section perpendicular to the flux in [m$^2$].

The second term ($\frac{\lambda}{\mu_{iron}}$) in Eq. 4 is called 'normalized reluctance' of the yoke.

If we are not careful enough with our design we can create more or less saturated areas in the iron yoke. Saturation means a local decrease of the iron permeability (small $\mu_{iron}$) which leads to inefficiencies of the magnetic circuit. It is good practice to keep the iron yoke reluctance smaller than a few per cent of air reluctance ($\frac{h}{\mu_0}$) by providing a sufficiently large iron cross-section such that the magnetic flux in the iron remains smaller than 1.5 T. If the recommendation ($\frac{\lambda}{\mu_{iron}} < 0.01\frac{h}{\mu_0}$) is followed diligently, the efficiency is better than 99%.

Efficiency: $$\eta = \frac{R_{M,\,gap}}{R_{m,gap} + R_{M,yoke}} \approx 99\ \%$$ .

### 4.1.6 Magnetic length

To understand the concept of magnetic length, we have to imagine approaching the magnet with a measurement probe along the beam axis from infinity towards the magnet centre. What we will read on the instrument is a steady increase of the field when we move closer to the edge of the iron yoke passing through the stray field of the magnet. The field continues to rise even when we are entering the gap of the magnet and will reach its maximum value when we move the probe towards the centre of the magnet where is remains more or less stable until we move again away from the centre towards the other end of the magnet. We see from Fig. 15 that the field does not increase suddenly but steadily when we approach the edge of the iron yoke. Integrating the magnetic field along the longitudinal axis starting from far outside on one side and ending far outside from the magnet on the other side will give a higher value than simply multiplying the local magnetic field with the iron length of the magnet. Here we can introduce the term 'magnetic length' $l_{mag}$ which is defined as

$$l_{mag} = \frac{\int_{-\infty}^{\infty} B(z) \cdot dz}{B_0}$$ . (7)

We can conclude that the magnetic length is always larger than the actual iron length. To calculate exactly the magnetic length analytically can be quite difficult. Usually it is derived from numerical computations by integrating the field along the magnet as described above and dividing it by the local field in the centre of the magnet. Nevertheless, there is a way to approximate the magnet length, which works well in cases where the iron length of the magnet is much larger than its gap height. For a dipole we can estimate the magnet length with

$$l_{mag} = l_{iron} + 2hk$$ (8)

where $k$ is a constant which is specific to the yoke geometry. The constant $k$ gets smaller when the pole width is smaller than the gap height, when saturation occurs in the pole regions, or when the coil heads are close to the beam axis. Typical values of $k$ are between 0.3 and 0.6. A precise determination of $k$ is only possible with measurements or numerical calculations.

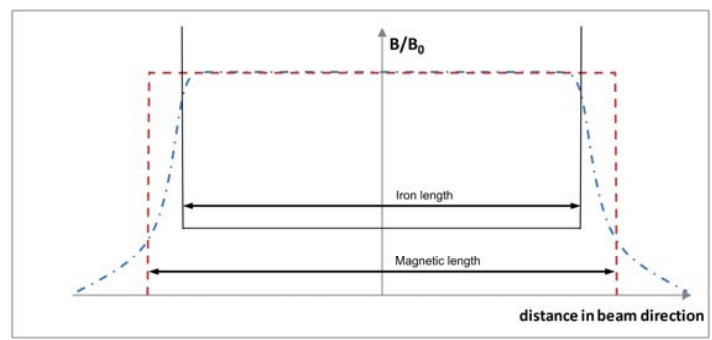

**Fig. 15:** Magnetic length – field distribution along the beam axis

### 4.1.7    *Magnetic flux*

The term magnetic flux Φ through a surface is defined as the integral of the normal component of the flux density over the cross-section area of this surface. In order to see whether there is any saturation issue in the iron we need to estimate the average flux density in the individual parts of the yoke.

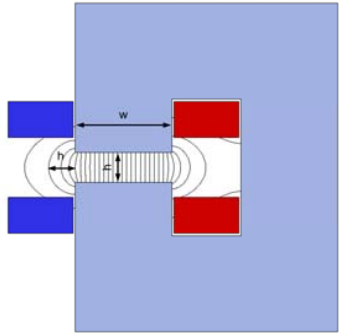

**Fig. 16:** Flux in the magnet aperture

This can be done by dividing the total magnetic flux by the cross-section area of the individual parts. As shown in Fig. 16 the flux entering the pole surfaces consists mainly of the useful flux in the gap, but for the correct calculation of the total flux in the return yoke we need to consider as well the stray flux entering on the sides of the poles. Again, a precise analytic calculation of the total flux is difficult, but for simple dipole geometry we can roughly estimate it by using the following relation.

Total flux in the return yoke is

$$\Phi = \int_A B \cdot da \approx B_{gap}(w + 2h)\, l_{mag} \tag{9}$$

where *h* is the gap height and *w* the pole width.

### 4.1.8    *Inductance*

To size the power converters, we need to know the maximum current and the RMS current, and the dc power consumption, but also the voltage that the converter has to supply. The total required voltage is a function of the maximum current to excite the coils, the resistance and inductance of the coils, and the envisaged speed to reach the maximum field. The total voltage on a ramped magnet is given by

$$V_{tot} = RI + L\frac{dI}{dt} \tag{10}$$

where *R* is the total electrical resistance of the excitations coils in [Ω, ohm], *L* is the total inductance of the magnet in [H, henry], and $\frac{dI}{dt}$ the maximum current ramp rate in [A/s].

For a magnet cycled with $I = I_0 \sin(\omega t)$ the total voltage is

$$V_{tot} = RI_0 \sin(\omega t + \varphi) \tag{11}$$

where $I_0$ is the amplitude of the current, ω is the angular frequency and $\varphi$ is the phase angle:

$$\varphi = tan^{-1}\frac{L\omega}{R} \,. \tag{12}$$

The coil resistance $R$ can be easily calculated taking into account the conductor length $l$, the effective cross-section $a$, and the resistivity $\rho$ of the conductor material:

$$R = \frac{l\rho}{a}.$$ (13)

This leads us to the question of how to calculate the inductance, which is not obvious. The inductance depends on the number of turns and on the coil geometry, but also on the geometry of the iron yoke surrounding the coils, which makes it more difficult to calculate correctly than for a simple cylindrical coil in free space. One possibility is to go via the stored energy $U$ [J, joule] in the magnet

$$L = \frac{2U}{I^2}$$ (14)

so that Eq. (10) becomes

$$V_{tot} = RI + L\frac{dI}{dt} = RI + \frac{2U}{I^2}\frac{dI}{dt}.$$ (15)

Unfortunately, this method has the drawback that we now have to compute the stored energy correctly, which is itself not easy. As the stored energy in a magnet depends on the non-uniform field distribution in the gap, the coils, and the iron yoke, it is usually determined by numerical computation.

However, for the very simple case of a window-frame magnet with constant field in the gap as shown in Fig. 17(c), the stored energy can be estimated as follows:

$$U_{gap} = \frac{B^2}{2\mu_0}V_{gap} \qquad\qquad U_{coil} = \frac{B^2}{6\mu_0}V_{coil} \qquad\qquad U_{yoke} = \frac{B^2}{2\mu_r\mu_0}V_{yoke}$$

where $V_{gap}$, $V_{coil}$, and $V_{yoke}$ are respectively the volumes of the gap, coil and yoke in [m$^3$]. Hence

$$U_{magnet} = U_{gap} + 2U_{coil} + U_{yoke} = \frac{B^2}{2\mu_0}\left(V_{gap} + 2\frac{V_{coil}}{6} + \frac{1}{\mu_r}V_{yoke}\right).$$

### 4.1.9 Dipole types

Although the design and layout of a dipole magnet can be quite different from case to case depending on the application, we can nevertheless identify three standard families typically used in particle accelerators and transfer lines: the so-called C-magnet, the H-magnet and the O- or window-frame magnet as shown in Fig. 17. Looking into their characteristics there is no optimum solution; they all have their advantages and drawbacks. The choice for one or the other option is led by the constraints and requirements such as the function of the magnet, the available space, and the field quality. The following sections provide a short overview of these three main types pointing out their pros and cons.

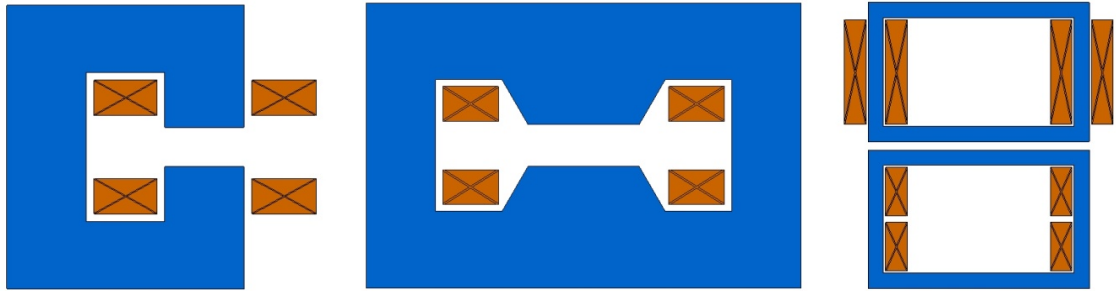

**Fig. 17:** Standard dipole types: C-magnet (a), H-magnet (b), and O-magnet (c)

### 4.1.9.1   C-magnet

The C-shaped magnet or C-magnet in Fig. 17 (a) provides a very good accessibility to the beam pipes, which makes it a perfect candidate for light sources where the synchrotron light has to be extracted all along the circumference of the synchrotron. Owing to its asymmetric layout this type is also suitable for injection and extraction regions or zones where adjacent beams are very close to each other like in the transfer lines of experimental areas.

The yoke volume and hence the weight of C-magnets is significantly higher than H-magnets with similar performance. The mechanical stability is less good compared to an H- or O-magnet since it has only one return leg and the attracting magnetic forces may lead to a movement of the poles when the magnet is pulsed. Transversal shims are usually required to achieve a decent field quality.

A drawback of the C-magnet is the asymmetrical field distribution in the gap. Unlike an H-magnet with two-fold symmetry around both axes, a C-magnet has only a one-fold symmetry. Because

$$NI = \oint \vec{H} \cdot \overline{dl}$$

has to be constant, the contribution to the integral in the iron has different path lengths, as shown in Fig. 18. The finite permeability will create lower field densities on the outside of the gap than on the inside which generates so-called 'forbidden' harmonics with n = 2, 4, 6, etc. Typically, the dipole produces a gradient across the pole of 0.1% with respect to the central field. In addition, the harmonics change with saturation and display non-linear behaviour depending on the excitation level.

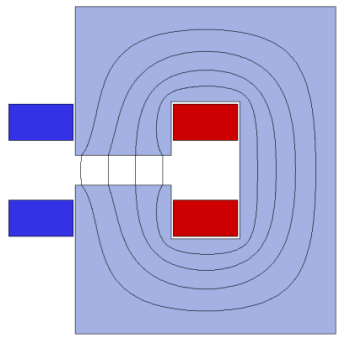

**Fig. 18:** Flux distribution in a C-magnet

### 4.1.9.2   H-magnet

The H-type magnet from Fig. 17(b) is used as standard in many accelerators and beam transfer lines. Access to the coils and beam pipes is poor, but they provide a good mechanical stability and a symmetric field quality. The iron weight is reduced with respect to C-magnets and they are usually made of two parts to allow an easy installation of the coils and the vacuum chamber. Transversal shims are also required here to achieve a decent field quality.

### 4.1.9.3 Window-frame or O-magnet

If we reduce the pole heights of an H-magnet to zero we basically arrive at the so-called window-frame layout. It has similar characteristics to the H-magnet in terms of symmetry, weight, and mechanical stability, with the difference that the window-frame design provides a very homogenous field quality even without shims. As shown in Fig. 17(c), there are two basic versions of this magnet type which employ different coil designs. The image on the bottom represents a classical window-frame magnet with saddle coils (see Section 4.5.1), while the version on the top uses racetrack coils installed around the vertical legs of the return yokes. The latter is less efficient in terms of excitation: it requires more ampere-turns compared to the version with the saddle coils. In addition, it generates a lot of stray field in the surroundings of the magnet. However, coils can also be installed around the horizontal leg of the magnetic circuit adding a vertical bending function. Such combined horizontal/vertical magnets are often used as steering magnets due to their compact design, but their efficiency is low.

## 4.2 Quadrupole yoke design

### 4.2.1 Magnetic induction

Analogous to the dipole, the required quadrupole field gradient $B'$ in [T/m] can be derived by using

$$B' = B\rho k \qquad (16)$$

where $k$ is the quadrupole strength in [m$^{-2}$].

### 4.2.2 Excitation current

The excitation current in a quadrupole can be calculated using similar considerations to those for a dipole. Choosing the integration path shown in Fig. 19 we get

$$NI = \oint \vec{H} \cdot \vec{dl} = \int_{s1} \vec{H_1} \cdot \vec{dl} + \int_{s2} \vec{H_2} \cdot \vec{dl} + \int_{s3} \vec{H_3} \cdot \vec{dl} \ .$$

For an ideal quadrupole, the gradient $B' = \dfrac{dB}{dr}$ is constant, $B_x = B'y$ and $B_y = B'x$ .

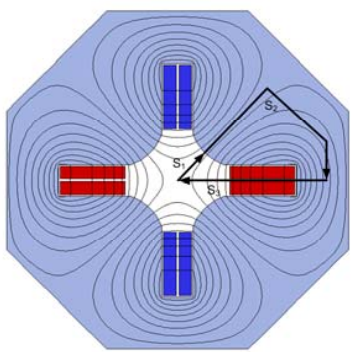

**Fig. 19:** Closed integration path in a quadrupole

The field modulus along the path $s_1$ can hence be written as

$$H(r) = \frac{B'}{\mu_0}\sqrt{x^2 + y^2} = \frac{B'}{\mu_0}r \quad .$$

Assuming that $\mu_{iron}$ is large we can neglect $B$ in the part $s_2$ because the reluctance $R_{M,s2} = \frac{s_2}{\mu_{iron}}$ in the iron is small compared to the reluctance in the air gap. Since $B_x$ on the $x$-axis ($y = 0$) is zero, the integral

$$\int_{s3} \vec{H}_3 \cdot \vec{dl} = 0$$

too, so we can also ignore the contribution of $B$ along path $s_3$.

This leads to

$$NI \approx \int_0^R H(r) \cdot dr = \frac{B'}{\mu_0}\int_0^R r \cdot dr$$

and finally to

$$NI_{(per\ pole)} = \frac{B'r^2}{2\eta\mu_0} \quad . \tag{17}$$

The highest magnetic field appears at the pole vertex.

### 4.2.3    *Magnetic length*

The magnetic length for a quadrupole can be estimated by

$$l_{mag} = l_{iron} + 2r\kappa \tag{18}$$

where $\kappa$ is a geometry specific constant (typically around 0.45) which can best be determined for a particular yoke geometry by numerical calculation.

It is interesting to note that the number of ampere-turns for a given gradient increases with the square of the quadrupole aperture and the dissipated power even with the power of four.

$$NI \propto r^2 \qquad\qquad\qquad P \propto r^4$$

This fact makes it more difficult to accommodate the necessary ampere-turns and coil cross-section in the iron yoke and to assure a sufficient cooling. To make space for the coil the hyperbola has to be truncated — digressing from the ideal pole profile. Depending on where the hyperbola is terminated, the resultant (allowed) higher order harmonics may affect the field quality in the aperture sufficiently to warrant correction.

### 4.2.4    *Quadrupole types*

In the same way as for the dipoles we can classify the different quadrupole layouts in several categories. The most common examples are presented in Fig. 20.

The standard quadrupole in Fig. 20(a) has 90° poles and provides very limited space for coils.

The image in Fig. 20(b) shows another standard quadrupole with parallel pole sides. It provides maximum space for coils, but the tendency to show saturation around the region of the pole roots limits operation as a high-field quadrupole. Note that the entire stray flux entering all along the pole side has to pass through the pole root. This design is used when moderate field gradients are required.

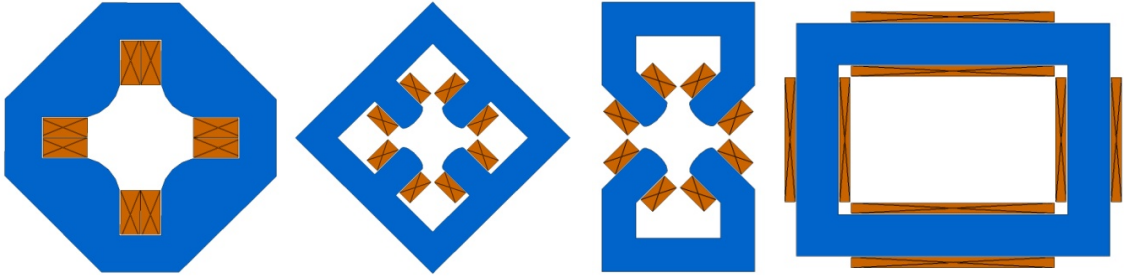

**Fig. 20:** Quadrupole types: standard type (a) and (b), Collins (c) and Panofsky (d)

A compromise between the two standard types is a quadrupole with tapered pole sides, which is not shown here. It combines the advantage of larger coil windows and less saturation in the pole roots, but is more complicated and costly to manufacture.

The so-called Collins or figure-of-eight quadrupole in Fig. 20(c) is a special type suitable for light sources and narrow beam lines because it provides an opening on the side allowing the extraction of beams or synchrotron light. It is obviously mechanically less stable, more complicated to produce, and therefore more expensive.

A more exotic type is the Panofsky quadrupole shown in Fig. 20(d). This type provides an excellent field quality; however it is used only as corrector magnet, because of its limited field strength. The Panofsky quadrupole looks like a window-frame dipole with horizontal and vertical coils, but is in reality an ironless magnet, since the magnetic field is determined by the current distribution in the copper conductors and not by the iron yoke.

## 4.3 Sextupole yoke design

### 4.3.1 Magnetic induction

Analogous to the dipole and the quadrupole, the differential field gradient $B''$ in [T/m$^2$] of a sextupole can be computed

$$B'' = B\rho m \tag{19}$$

with $m$ being the sextupole strength in [m$^{-3}$].

### 4.3.2 Excitation current

To identify the required number of ampere-turns for a sextupole we chose the same approach as for quadrupoles. For a sextupole, the field is parabolic and $B'' = \dfrac{d^2 B}{dr^2}$ is constant so that

$$H(r) = \frac{B''}{2\mu_0} r^2$$

leading to

$$NI = \oint \vec{H} \cdot \vec{dl} \approx \int_0^R H(r) \cdot dr = \frac{B''}{2\mu_0} \int_0^R r^2 \cdot dr$$

and

$$NI_{(per\ pole)} = \frac{B'' r^3}{6\eta\mu_0}.$$

(20)

Analogous to the quadrupole, the ampere-turns in the sextupole increase with the third power of the aperture and the power dissipated in the coils rises with the sixth power of the aperture.

$$NI \propto r^3 \qquad\qquad P \propto r^6$$

Fortunately, sextupolar fields are usually required to be much smaller than quadrupole fields.

### 4.4   Yoke materials

Magnetic circuits or yokes are made of magnetic steel. They can be built from massive iron or assembled from laminations. Historically, the primary choice for either of these techniques was whether the magnet was cycled or operated in persistent mode. Solid yokes support eddy currents and hence cannot be cycled or pulsed rapidly. To reduce or avoid eddy currents in pulsed operation the yoke has to be laminated.

Yokes machined from cast ingots require less tooling than for the stamping, stacking and assembling of laminated yokes. A major problem with massive yokes is the difficulty of providing magnets with similar magnetic performance. To assure identical characteristics within the accepted tolerances all yokes have to be built using the same melt. This requires very careful documentation.

Today's practice— even for dc operated magnets — is to use cold-rolled, non-grain-oriented (NGO) electro-steel sheets and strips (according to EN 10106). Although laminated yokes are labour intensive and require more and expensive tooling they offer a number of advantages:

–   Magnetic and mechanical properties can be adjusted by final annealing

–   Reproducible steel quality even over large productions

–   Magnetic properties (permeability, coercivity) within small tolerances

–   Homogeneity and reproducibility among the magnets of a series can be enhanced by selection, sorting or shuffling of the laminations according to their magnetic properties

–   Organic or inorganic coating for insulation and bonding

–   Material is usually cheaper

Table 2 summarizes typical material properties of cold-rolled, non-grain-oriented electro-steel. More detailed information on specific materials can be requested from the steel producers.

**Table 2**: Typical properties of cold-rolled, non-grain-oriented electro-steel

| Property | Typical value |
| --- | --- |
| Sheet thickness | $0.3 \leq t \leq 1.5$ mm |
| Density | $7.60 \leq \delta \leq 7.85$ kg/dm$^3$ |
| Coercivity | $H_c < 65$ A/m |
| Coercivity spread | $\Delta H_c < \pm 10$ A/m |
| Electrical resistivity at 20°C | $0.16$ (low Si) $\leq \rho \leq 0.61$ μΩm (high Si) |

### 4.4.1 Permeability

At the beginning of this section we discussed the relation between the magnetic field strength $H$ and the magnetic flux density $B$, which is defined for free space as

$$B = \mu_0 H \quad ,\tag{21}$$

where $\mu_0$ is a universal constant with the value $4\pi \cdot 10^{-7}$ Vs/Am. For the magnetic flux density in a material Eq. (21) becomes

$$B = \mu H$$

where $\mu$ is called the absolute permeability which is material specific. Because

$$\mu = \mu_0$$

for free space, we can relate the permeability of matter to the permeability of free space by

$$\mu = \mu_r \mu_0 \tag{22}$$

introducing $\mu_r$ as the dimensionless relative permeability, which characterizes the magnetic behaviour of materials. We can distinguish between three main categories of materials:

  – diamagnetic materials ($\mu_r < 1$),

  – paramagnetic materials ($\mu_r > 1$), and

  – ferromagnetic materials ($\mu_r \gg 1$).

For the construction of electromagnets, only the third category is important. Several examples of commonly used steel grades and their relative permeability are presented in Fig. 21.

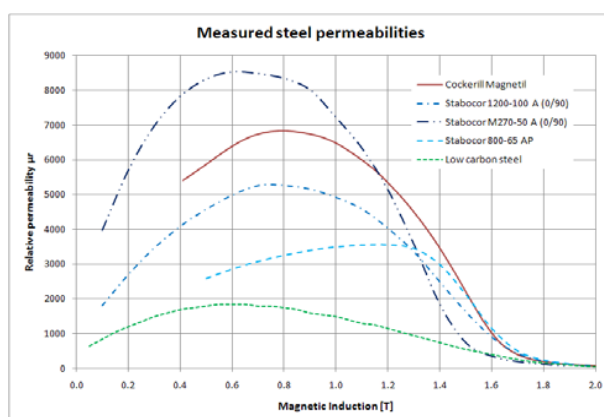

**Fig. 21:** Permeability of different steel grades

### 4.4.2 Magnetic polarization

Ferromagnetic materials show a non-linear correlation between the field strength $H$ and the flux density $B$. The total flux density in the material is the sum of the flux density in free space $\mu_0 H$ and the magnetic polarization $J$ in [T] and is described by the equation

$$B = \mu_0 H + J = \mu_r \mu_0 H \quad .\tag{23}$$

The magnetic polarization *J* for specific materials is typically presented in tables or graphs by measured data. In addition to the non-linear behaviour, cold- and hot-rolled steel have material characteristics which can be anisotropic to a high degree. This anisotropy, in particular for the permeability can be partly cured by final annealing, but remains to a certain extent and cannot be neglected, so it has to be considered in the magnetic design. Fig. **22** shows the anisotropic polarization and permeability of cold-rolled electro-steel (grade 1200 – 100A) after final annealing.

### 4.4.3    *Hysteresis, remanence and coercivity*

On account of complicated material internal processes (movements and growth of magnetic domains) we can observe a hysteresis, which means that the flux density *B(H)* as a function of the field strength is different when increasing and decreasing excitation. This behaviour is shown in Fig. 23.

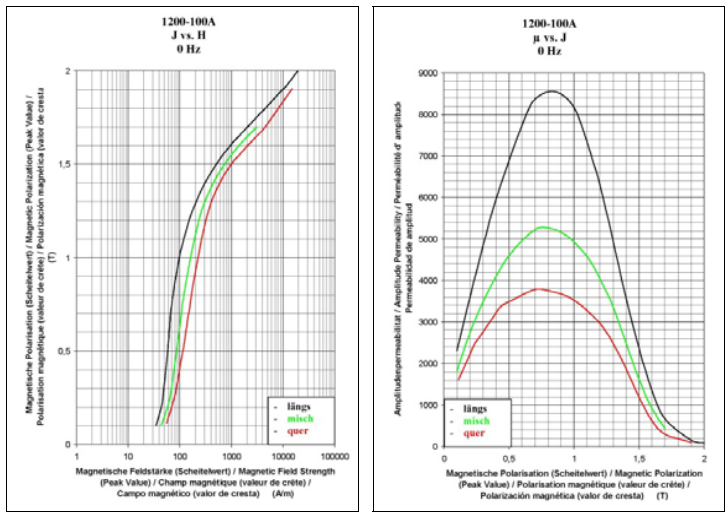

**Fig. 22:** Anisotropic polarization (a) and permeability (b); data source: Thyssen/Germany

When the current is switched off, some magnetic polarization of the iron remains: this is called remanent field or magnetic remanence $B_r$. The width of the hysteresis curve is determined by the coercive force or coercivity $H_c$. The quantity $H_c$ is defined as the value of field strength that reduces the magnetic flux density in the steel to zero. Materials having $H_c < 1000$ A/m are called soft magnetic materials, e.g., electro-steel, those with $H_c > 1000$ A/m are called hard magnetic, e.g., permanent magnets.

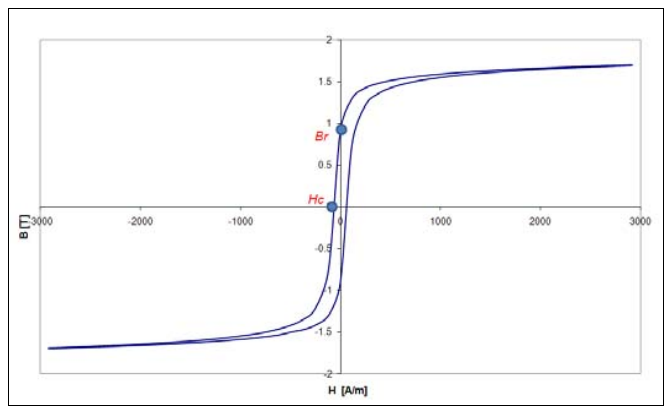

**Fig. 23:** Hysteresis curve of electro-steel (grade 1200 – 100 A)

In a continuous ferromagnetic core, as is the case of a transformer, the residual field is entirely determined by the remanence $B_r$. In a magnet where the highest reluctance appears in the magnet gap, the residual field is mainly determined by $H_c$. If we take a magnet which has been excited to a certain field level and we switch off the current in the coils we get

$$NI = \oint \vec{H} \cdot \overrightarrow{dl} = \int_{gap} \overrightarrow{H_{gap}} \cdot \overrightarrow{dl} + \int_{yoke} \overrightarrow{H_c} \cdot \overrightarrow{dl} = \mathbf{0} \tag{24}$$

and so

$$B_{residual} = -\mu_0 H_c \frac{\lambda}{g} \tag{25}$$

To set the residual field to zero, a negative current must be sent through the coils. In practice it is often more convenient to set a zero field in the magnet by running it through a certain number of so-called demagnetization cycles. In normal operation, the magnet is always cycled to its maximum value, irrespective of the required field, to ensure that hysteresis effects are reproducible.

## 4.5   Coil design

In the previous sections it has been shown how to determine the necessary ampere-turns. In this part we will see how to choose a current density, the number of turns and dimensions of the coil.

The design of the coils is not completely independent from the layout of the yoke. Optimizing the coils, e.g., for low power consumption, is usually at the expense of a larger yoke cross-section. It is the duty of the magnet designer to find the right compromise between a good coil design and a good yoke design. A high-quality coil design unifies low electrical power consumption, sufficient cooling performance, adequate insulation thickness, and moderate material and manufacturing costs. To reach this goal, and to achieve a satisfactory overall magnet performance, requires several iterations.

The coil design sequence can be divided into steps:

- Selecting an adequate coil type
- Calculating power requirements
- Cooling circuit configuration
- Selecting the conductor dimensions
- Optimization

### 4.5.1   Standard coil types

As in the case of the yoke, coil layouts can also differ, depending on the application. Fig. **24** illustrates the types most frequently used for normal-conducting accelerator magnets.

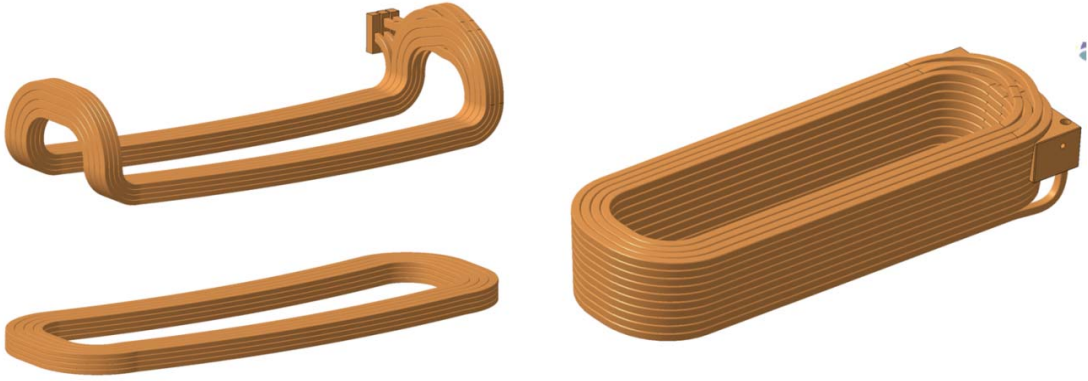

**Fig. 24:** Standard coil types: bedstead coil (a), racetrack coil (b) and quadrupole coil (c)

*Racetrack coils:* This coil type is relatively easy to manufacture and therefore the least expensive. It is commonly used in magnets with poles (C-magnet, H-magnet). To facilitate production and installation, the coils are often manufactured in pancakes, in particular for C-shape magnets.

*Bedstead or saddle coils:* They are more complicated to wind and require a complex impregnation mould, which makes them more expensive. This type is used for O-type and H-type magnets. The bent coil heads allow filling the entire coil window with conductor material and leaving space for the beam pipes and the magnet ends.

*Quadrupole coils:* This coil can be used in quadrupoles with parallel or slightly tapered poles. A particularity shown here are the integrated terminals for water and electricity.

### 4.5.2 Power requirements

Once the coil type has been selected we need to determine the power requirements. For this we assume that the magnet cross-section and the yoke length are known. The total dissipated power for the individual magnet types can be calculated accordingly:

dipole:
$$P_{dipole} = \rho \frac{Bh}{\eta \mu_0} j l_{avg} \tag{26}$$

quadrupole:
$$P_{quadrupole} = 2\rho \frac{B' r^2}{\eta \mu_0} j l_{avg} \tag{27}$$

sextupole:
$$P_{sextupole} = \rho \frac{B'' r^3}{\eta \mu_0} j l_{avg} \tag{28}$$

where $\rho$ is the resistivity in [$\Omega$m], $l_{avg}$ is the average turn length in [m] (a useful approximation is 2.5 $l_{iron} < l_{avg} < 3\ l_{iron}$ for racetrack coils).

The current density $j$ in [A/m$^2$] is defined as

$$j = \frac{NI}{f_c A} = \frac{I}{a_{cond}} \tag{29}$$

with $a_{cond}$ being the conductor cross section in [m$^2$], $A$ being the coil cross section in [m$^2$] and $f_c$ being a dimensionless geometric filling factor ($= \dfrac{net\ conductor\ area}{coil\ cross\ section}$) taking into account insulation material, cooling duct and the conductor edge rounding. It is interesting to note that for a constant geometry, the power loss $P$ is proportional to the current density $j$.

### 4.5.3   Number of turns

The power which we have determined above can be divided into a voltage and a current according to

$$P = VI$$

With the help of the following basic relations

$$R_{magnet} \propto N^2 j \qquad\qquad V_{magnet} \propto Nj \qquad\qquad P_{magnet} \propto j$$

we can choose a number of turns $N$ to match the impedances of the power converter.

A large number of turns implies low current but high voltage which consequently requires thicker insulation for both coils and cables, which gives rise to a poor filling factor. A positive effect is that the transmission power losses are kept low even across long distances between the power converter and the magnets. The choice of coils with many turns is therefore made primarily for magnets with moderate magnetic field strength which are powered individually.

A small number of turns implies high current but low voltage. The drawbacks are large terminals and conductor cross-section. Advantages are a better conductor filling factor in the coils, smaller coil cross-sections and less stringent demands on the coil and cable insulation. Since the transmission power losses are high, this solution is chosen when many magnets have to be electrically connected in series and the distance from one magnet to the next is relatively short, such as in the case of bending magnets in a synchrotron. In this case to have many turns would lead to unreasonably high voltages between the coils and the magnet yokes increasing the risk of short circuits. The transmission power loss can be handled by using water-cooled cables or rigid bus bars with large cross-sections.

## 4.6   Cooling

The electrical power which is dissipated in the coils has to be removed from the magnets otherwise overheating can seriously damage the coil insulation and cause short circuits between the coil conductor and the surrounding equipment which is usually on ground potential. In the field of normal-conducting magnets, we distinguish between two different cooling techniques: air cooling and water cooling. Sometimes they are also referred to as 'dry cooling' and 'wet cooling'.

### 4.6.1   Air cooling

Air cooling by natural convection is suitable only for low current densities. This limits the application to magnets with moderate field strength like correctors or steering magnets. As a rule of thumb, the maximum current density for voluminous coils which are almost entirely enclosed in the magnet yoke should not exceed 1 A/mm$^2$. For small, thin coils, current density can be higher, but below 2 A/mm$^2$.

A precise thermal study of air-cooled magnets by analytic means is difficult if not impossible. Air cooling is a combination of convection, radiation and heat conduction and depends on coil geometry, coil surface (roughness, material, colour), thermal contact to the surrounding materials, etc. Detailed analysis of the thermal behaviour, if needed, would require numerical computations or measurements. Information on air cooling and cooling in general can be found in the relevant text books in the bibliography at the end of this paper.

Air-cooled coils can be made of round, rectangular, or square wires. They are commercially available in various grades and dimensions and can be ordered blank or pre-impregnated with varnish (0.02 $\leq$ t $\leq$ 0.1 mm) or half-overlapped polyimide (Kapton®) tape (0.1 $\leq$ t $\leq$ 0.2 mm). Depending on the winding precision, the insulation thickness, and the conductor cross-sections a filling factor between 0.63 (round) to 0.8 (rectangular) can be obtained. The outer or ground insulation is typically made by epoxy impregnated glass fibre tapes of thickness between 0.5 mm and 2 mm.

The cooling performance of air-cooled coils can be enhanced by mounting an appropriate heat sink with enlarged radiation surface or by forced air flow (cooling fan).

### 4.6.2    Water cooling

There are two methods of water or wet cooling: direct and indirect. The latter is of minor importance and rarely used, although it has the advantage that normal tap water can be used as coolant, which does not require cooling plants with water treatment for the supply of demineralized water. In the absence of such infrastructure, indirect cooling should be considered as a possible alternative, even though it implies a more complex coil design. In a situation where air cooling is just at the limit, the mounting of an external heat sink cooled with tap water can enhance the cooling performance keeping the thermal load within limits or permitting slightly higher current densities. However, indirect cooling is seldom used, so here we focus on the engineering and construction of direct water-cooled coils.

The current density in direct water-cooled coils can be typically as high as 10 A/mm$^2$. This is a conservative value that can be easily realized with standard coil models. It is a good compromise assuring a high level of reliability during operation and a compact coil layout. Although current densities of 80 A/mm$^2$ can be attained for specific applications, e.g., septum magnets, it is not recommended for standard magnets because the reliability and lifetime of the coils is significantly reduced. High current densities require a sophisticated cooling circuit design with multiple parallel circuits per coil — even single turn cooling — and high coolant velocities increasing the risk of erosion.

Standard water-cooled coils are wound from rectangular or square copper or aluminium conductor with a central cooling duct for demineralized water as shown in Fig. 25. The inter-turn and ground insulation is provided by one or more layers of half-lapped glass fibre tape impregnated in epoxy resin. Inter-turn insulation thickness is normally between 0.3 mm and 1.0 mm, the ground insulation thickness should be between 0.5 mm and 3.0 mm depending on the applied voltage.

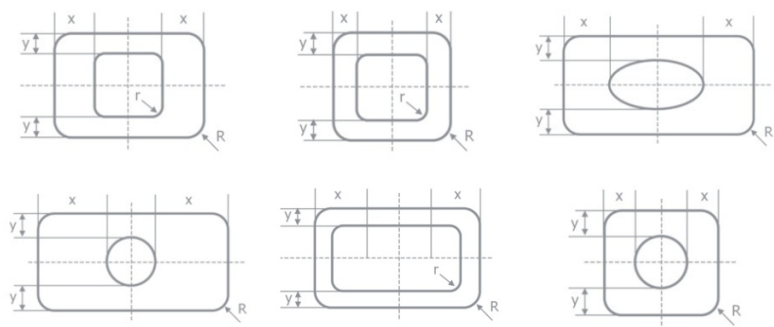

**Fig. 25:** Hollow conductor profiles for water-cooled coils

### 4.6.3    Conductor materials

Conductor materials which are most commonly used for the construction of normal-conducting are aluminium and copper. Both materials are available in different grades and purities. While aluminium was a good alternative in the past, since it was much cheaper, nowadays copper coils are more state-of-the-art. The material costs today are similar for a copper coil and an aluminium coil provided they have the same electrical resistance, but aluminium coils have a larger cross-section. On the other hand, aluminium is easier to form since it does not work harden.

Table 3 gives two examples of typical standard materials together with their main characteristics. Magnet designers and engineers who look for more detailed information and other available grades should consult suppliers, material databases, or international standards.

**Table 3**: Typical properties of conductor materials

| Property | Aluminium | Copper (OF grade) |
|---|---|---|
| Purity | 99.7% | 99.95% |
| Resistivity at 20°C | 28.3 nΩ m | 17.2 nΩ m |
| Thermal resistivity coefficient | 0.004 K⁻¹ | 0.004 K⁻¹ |
| Density | 2.70 kg/dm³ | 8.94 kg/dm³ |
| Thermal conductivity | 237 W/m K | 391 W/m K |
| Approx. price | 4.7 €/kg | 11 €/kg |

When winding a coil — regardless of whether it is made of aluminium or copper — special care has to be taken to avoid small bending radii. A tight bending radius increases the risk of insulation damage, decreases the cross-section of the cooling duct and leads to an increase of the outer conductor dimensions, which is known as the 'keystone effect' or 'keystoning'.

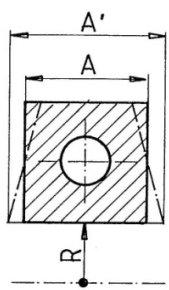

**Fig. 26:** Keystone effect

Keystoning, as shown in Fig. 26, is caused by the horizontal growth of the conductor on the inside radius of the coil when bending the conductor around the horizontal axis. The effect of keystoning has been determined in the past empirically. For a bending radius of three times the conductor width we can expect a keystoning of 3.6%:

$$R = 3 \cdot A \Rightarrow \frac{\Delta A}{A} = 3.6\%$$

where $A$ means the conductor dimension parallel to the bending axis.

For coils with many turns, the cumulative effect of keystoning can lead to a significant enlargement of the coil dimensions in the bending regions and consequently to problems for the installation of the coils in the yoke. In principle, we can ignore the effect of keystoning by systematically choosing a bending radius four times larger than the conductor width.

### 4.6.4 Cooling parameters

Water cooling is necessary whenever large coil cross-sections have to be avoided and higher current densities are indispensible. The construction of water-cooled coils requires hollow conductors with a cooling channel. Choosing the correct parameters and dimensions, such as number of cooling circuits, the size of the cooling channel and the flow rate, is not easy and requires several iterations to arrive at a satisfactory solution. This section addresses the calculation and design of an efficient cooling circuit.

Before entering into the subject, I will give a few recommendations and canonical values, which should help the magnet designer with his choices. Please note that these values are rules of thumb and should be critically cross-checked as to their validity and applicability in the particular situation.

As mentioned in the previous section, the current densities for water-cooled coils should be kept between 2 A/mm² and 10 A/mm². Below these values, air cooling is probably sufficient; exceeding this threshold implies more complicated and costly designs.

The pressure drop which is typically provided by modern cooling plants ranges between 0.1 MPa and 1.0 MPa, which corresponds to 1 to 10 bar. Advanced cooling stations can supply water with a pressure up to 2.0 MPa. A low pressure drop can be compensated to a certain extent by a higher cooling flow and a more complex (and expensive) coil design with several cooling circuits in parallel.

The velocity of the cooling medium — in general demineralized water — should be sufficiently high to guarantee a turbulent flow but low enough ($u_{avg} \leq 5$ m/s) to avoid erosion and vibration.

A maximum permitted temperature of less than 60°C on the coil surfaces was found to be good practice. It avoids or reduces accelerated ageing of insulation materials in particular in presence of ionizing radiation. Assuming a water inlet temperature of less than 30°C, this corresponds to a maximum water temperature rise of 30°C, also keeping thermal stress within limits. For some applications a high mechanical stability and consequently a good thermal stability is essential. In such cases, the maximum temperature rise has to be reduced accordingly.

In what follows it is assumed that the cooling pipes are always long, mostly straight and smooth inside without perturbations and that the flow is turbulent, meaning a high Reynolds number. We also assume good heat transfer from conductor to coolant, with the temperature of the inner surface of the conductor equal to that of the coolant, and that the conductor is isothermal over its cross-section.

The first step is to determine how the velocity of a coolant depends on the dimensions of the cooling pipe and the pressure drop along it. The pressure drop through a water circuit is known to be

$$\Delta p = f \frac{l}{d} \frac{\delta u_{avg}^2}{2}$$

(30)

with $p$ being the pressure in [Pa, N/m²], $f$ a dimensionless friction factor, $l$ and $d$ the cooling circuit length in [m] and diameter in [m], $\delta$ the coolant mass density in [kg/m³], which for water is 1000 kg/m³ = 1 kg/litre, and $u_{avg}$ the average coolant velocity in [m/s].

The friction factor $f$ depends on the Reynolds number Re

$$\mathrm{Re} = \frac{u_{avg} d}{v}$$

(31)

with $v$ describing the temperature-dependent kinematic viscosity of a coolant. For our purpose we can assume it as constant: $9.85 \cdot 10^{-7}$ m²/s for water at 21°C.

From the text books, the friction factor $f$ for laminar flow (Reynolds number less than 2000) is

$$f = \frac{64}{\mathrm{Re}} .$$

For our purpose, the flow is turbulent (Re > 4000), and the friction factor $f$ is transcendental:

$$\frac{1}{\sqrt{f}} = -2 log_{10} \left( \frac{\varepsilon}{3.7 d} + \frac{2.51}{\mathrm{Re}\sqrt{f}} \right)$$

(32)

where $\varepsilon$ is a measure for the roughness of the cooling channel (typically $1.5 \cdot 10^{-3}$ mm). We can reformulation Eq. (30) to express the coolant velocity $u_{avg}$

$$u_{avg} = \sqrt{\frac{2 \Delta p \, d}{\delta \, f \, l}}.$$

(33)

The friction factor $f$ in Eq. (33) depends on the Re$(u_{avg})$, which means that to solve for $u_{avg}$ requires an iterative process. This is also shown in text books on heat transfer. Substituting the Reynolds number in Eq. (32) by

$$Re = \frac{d}{v} \sqrt{\frac{2 \Delta p \, d}{\delta \, f \, l}}$$

(34)

yields

$$u_{avg} = -2 \sqrt{\frac{2 \Delta p \, d}{\delta \, f \, l}} \, log_{10} \left( \frac{\varepsilon}{3.7d} + \frac{2.51}{\frac{d}{v} \sqrt{\frac{2 \Delta p \, d}{\delta \, f \, l}}} \right).$$

(35)

Assuming that water is used as cooling fluid, this complicated expression can be simplified to

$$u_{avg} \approx 0.3926 \cdot d^{0.714} \left( \frac{\Delta p}{l} \right)^{0.571}.$$

(36)

The heat absorbed by coolant medium across a heated surface is

$$P = \dot{m} c_p \Delta T$$

(37)

where $P$ is the power dissipated in [W], $c_p$ is the heat capacity in [W s/kg °C], which is $4.19 \cdot 10^3$ W s/kg °C for water, $\Delta T$ is the temperature increase in [°C] and $\dot{m}$ is the mass flow in [kg/s]. Writing

$$\dot{m} = \delta Q$$

(38)

with $Q$ being the coolant flow rate in [litre/s], the expression for the necessary flow $Q$ to remove the dissipated heat $P$ by the selection of the maximum permissible temperature $\Delta T$ is given by

$$Q = \frac{P}{\delta \, c_p \Delta T}.$$

(39)

Using water as cooling medium, Eq. (39) can be simplified and rewritten to

$$Q_{water} = 2.388 \cdot 10^{-4} \frac{P}{\Delta T}.$$

(40)

Knowing that the coolant flow inside a round tube with a bore diameter $d$ is

$$Q = u_{avg} \frac{\pi d^2}{4} 10^3,$$

(41)

the temperature increase $\Delta T$ in a water-cooled circuit due to dissipated power $P$ is

$$\Delta T = 3.04 \cdot 10^{-7} \frac{P}{u_{avg} \, d^2}$$

$\qquad\qquad$ (42)

Two further relations affect cooling performance and should be considered in the system layout:

Pressure drop is inversely proportional to the third power of the number of circuits $K_w$ per coil:

$$\Delta p \propto \frac{1}{K_w^3}$$

This implies that for a given flow, the pressure drop is reduced by a factor of eight by doubling the number of cooling circuits.

The pressure drop is inversely proportional to the fifth power of the cooling channel diameter $d$:

$$\Delta p \propto \frac{1}{d^5}$$

This implies that an increase of the cooling channel diameter by a small factor can reduce the required pressure drop significantly.

### 4.6.5    Cooling circuit design

In the previous section we derived the formulas required to calculate the main parameters of a cooling circuit. Finding a good solution for a cooling layout is an iterative and lengthy exercise and there are certainly different methods which lead to a satisfactory design. In this section I will demonstrate a way to start with the design of a cooling circuit for a dipole magnet, which has always worked fine for me. Other magnet types such as quadrupoles or sextupoles can be treated in a similar way by modifying the parameters accordingly. The steps described contain several approximations and assumptions which may not appear very precise, but they should help setting a starting point for an iterative design process. Depending on how demanding the requirements and constraints are, more or less fine-tuning on initial assumptions will be required in a second or third iteration. To start with we suppose that current density $j$, power $P$, current $I$, and the number of turns $N$ have already been determined.

First, the number of layers $m$ in the coil and the number of turns per layer $n$ has to be selected. Often one needs to round up the number of turns $N$ per coil to get reasonable numbers for $m$ and $n$. The coil height $c$ and coil width $b$ can be defined more or less arbitrarily, the limiting factor being the available space in the coil window of the yoke. An aspect ratio of $c{:}b$ between 1:1 and 1:2 should be chosen, and the packing factor $f_c$ somewhere between 0.6 and 0.8. The total coil cross-section area $A$ is

$$A = b \, c = \frac{N \, I}{j \, f_c}$$

$\qquad\qquad$ (43)

The following formula has been found to be a good approximation to estimate the average length of a conductor turn for dipole magnets with poles:

$l_{avg}$= pole perimeter + 8 × clearance between pole and coil + 4 × coil width

For the first iteration we can assume a coil with a single cooling circuit. If it turns out to be insufficient, the number of circuits per coil has to be increased in the second iteration. On the other hand, one may find that the required water pressure is far less than the available pressure. In this case one can simplify the magnet design by connecting several coils in series or reducing the size of the cooling duct, such that more conductor cross-section is available to carry the electrical current.

The total length of the cooling circuit, assuming that we can neglect the connection hoses or tubes since they are short compared to the circuit length in the coils, is expressed by

$$l = \frac{K_c \, N \, l_{avg}}{K_w}$$

(44)

where $K_c$ is the number of coils per magnet and $K_w$ is the number of water circuits per magnet. In the case of our former assumption to have one circuit per coil, $K_w$ is equal to $K_c$.

The next step is to select the maximum permissible temperature rise $\Delta T$ and the available pressure drop $\Delta p$ and to calculate the cooling hole diameter $d$

$$d = 5.59 \cdot 10^{-2} \left( \frac{P}{\Delta T \, K_w} \right)^{0.368} \left( \frac{l}{\Delta p} \right)^{0.21}.$$

(45)

If the results from Eq. (43) lead to a diameter of the cooling hole that is too large or to small, either the pressure drop $\Delta p$ or the number of cooling circuits $K_w$ can be changed before repeating the calculation. Once a satisfactory solution is found one can continue to determine the conductor area by

$$a = \frac{I}{j} + \frac{d^2 \pi}{4} + r_{edge}(4 - \pi)$$

(46)

where $r_{edge}$ is the edge rounding radius of the conductor profile. Knowing the conductor area $a$, the conductor dimensions and insulation thickness can be fixed. Now it is time to verify if the resulting coil dimensions, the number of turns $N$, the coil current $I$, the coils resistance $R$ and the temperature rise $\Delta T$ are still compatible with the initial assumptions and requirements. If this is not the case, another iteration with a new set of parameters has to be launched.

The last step is to compute the coolant velocity and the coolant flow using Eq. (36) and Eq. (40) and verify, using Eq. (31), whether the Reynolds number corresponds to turbulent flow (Re > 4000).

### 4.6.6   Cooling water properties

For the cooling of hollow conductor coils, with the exception of indirect cooled coils, demineralized water has to be made available. Since the water quality is essential for the performance and the reliability of the coil, the following typical water properties should be guaranteed:

- Water resistivity higher than $0.1 \cdot 10^6$ $\Omega$m
- pH-value between 6 and 6.5
- dissolved oxygen below 0.1 ppm (parts per million)

In addition to filters to be installed in the water circuits close to the magnet to avoid an obstruction of the cooling duct by particles, loose deposits and grease, a constant monitoring of the water quality is indispensible. A poor water quality neglecting the above recommendations might lead sooner or later to failures like electrical short circuits and leaks in the water circuit due to corrosion and erosion effects.

### 4.7  Cost estimate

Preparing a correct cost estimate for a magnet project requires a lot of experience and a good knowledge of the market. While the material prices are well known, the evaluation of the costs of manufacturing a magnet is not a precise science and depends on many different parameters. Past experience has shown that prices from the lowest to the highest bid may vary by up to a factor three.

In the framework of these lectures we can only give an order of magnitude indication of prices, as shown in Table 4. The inexperienced magnet designer should consult either more experienced colleagues and/or magnet suppliers to draft realistic estimation of costs.

**Table 4**: Cost indication for standard magnets (valid for 2010)

| Item | Cost indication |
| --- | --- |
| Production-specific tooling | $5000 - 15\,000$ €/tooling |
| Steel sheets | $1.0 - 1.5$ €/kg |
| Copper conductor | $10 - 15$ €/kg |
| | |
| *Yoke manufacture:* | |
| Dipoles (> 1000 kg) | $6 - 10$ €/kg |
| Quadrupoles, sextupoles (> 200 kg) | $50 - 80$ €/kg |
| Small magnets | up to 300 €/kg |
| | |
| *Coil manufacture:* | |
| Dipoles (> 200 kg) | $30 - 50$ €/kg |
| Quadrupoles, sextupoles (> 30 kg) | $65 - 80$ €/kg |
| Small magnets | up to 300 €/kg |
| | |
| Contingency | $10 - 20\%$ |

#### 4.7.1  Cost optimization

One of the main goals in magnet design should be to make an economic design for minimum total cost over the magnet lifetime. Total cost can be divided into capital investment and running costs as shown in Fig. 27. The cost factor 'power' is important as it enters into both the capital and running costs.

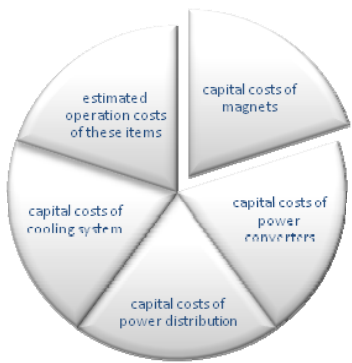

**Fig. 27:** Magnet cost factors

While it is difficult to estimate the absolute magnet costs precisely, the relative cost of design variants can be compared and optimized relatively easily. An interesting method to minimize overall cost by optimizing scaling parameters, e.g., magnet length and the current density is described in Ref. [2].

As demonstrated in this paper, a decrease of the current density leads to increases of the coil cross-section and increases of the coil and yoke material and manufacturing cost whilst decreasing the operation costs. The advantages of low current densities are the lower power losses, reducing the power consumption and power converter size and leading to less heat dissipated into the machine tunnel. High current density gives the advantages of smaller coils, smaller magnets, and lower investment cost.

## 5    Conclusions

I have shown in this paper how to start a magnet design by using simple analytical and empirical formulas. Such a first draft is often sufficient to validate the feasibility of a design proposal and to derive a list of the main magnet parameters. In addition it provides a solid foundation for the following steps where this preliminary design has to be refined by means of numerical programs. With the help of two- and three-dimensional FEM models one has to look in more detail into subjects like end field effects, field homogeneity and possible transient effects due to time variant parameters before ending with a final optimization.

### Acknowledgements

I wish to thank the CAS organizing committee for giving me the opportunity to present these lectures, which was a wonderful experience for me. In preparing these lectures and the notes I have profited greatly from the books and articles quoted in the bibliography, in particular from those of N. Marks in the CAS Proceedings, J. Tanabe and G. E. Fischer. Special thanks go to T. Taylor who was kind enough to review this paper by making valuable suggestions. Finally, I would like to thank my colleagues F. Luiz and C. Siedler for helping me to produce the nice figures which make the text more lively and comprehensible.

**Appendix A: Basic magnet types**

| Magnet type | Field distribution | Pole equation | Flux density ($B_y$) |
|---|---|---|---|
| 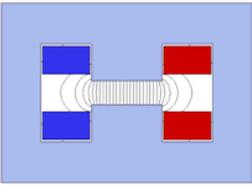 | 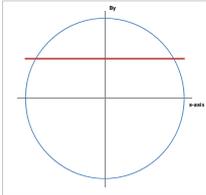 | $y = \pm r$ | $B_y = a_1 = B_0 = $ const. |
| 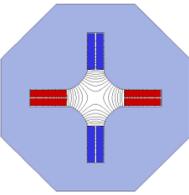 | 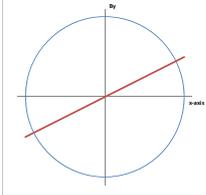 | $2xy = \pm r^2$ | $B_y = a_2 x$ |
| 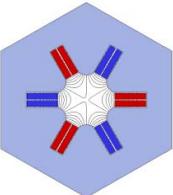 | 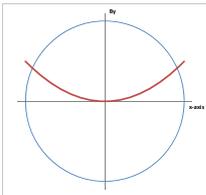 | $3x^2y - y^3 = \pm r^3$ | $B_y = a_3(x^2 - y^2)$ |
| 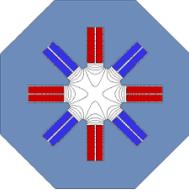 | 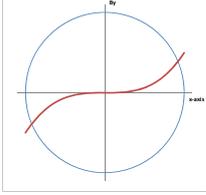 | $4(x^3y - xy^3) = \pm r^4$ | $B_y = a_4(x^3 - 3xy^2)$ |